\documentclass[sigconf]{acmart}
\usepackage{algorithm,float}
\usepackage{lipsum}
\usepackage{xcolor}
\usepackage{soulutf8}
\usepackage{empheq}
\usepackage{algorithmicx}
\usepackage{algpseudocode}
\usepackage{amsmath}
\usepackage{graphicx}
\usepackage{microtype}
\usepackage{nicefrac}
\usepackage{framed}
\usepackage{color,soul}

\AtBeginDocument{%
  \providecommand\BibTeX{{%
    \normalfont B\kern-0.5em{\scshape i\kern-0.25em b}\kern-0.8em\TeX}}}
\setcopyright{acmlicensed}
\copyrightyear{2018}
\acmYear{2018}
\acmDOI{XXXXXXX.XXXXXXX}

\acmConference[Conference acronym 'XX]{Make sure to enter the correct
  conference title from your rights confirmation emai}{June 03--05,
  2018}{Woodstock, NY}
\acmISBN{978-1-4503-XXXX-X/18/06}

\begin{document}
\makeatletter
\def\@copyrightspace{\relax}
\makeatother

\title{Balancing SoC in Battery Cells using Safe Action Perturbations}





\author{E Harshith Kumar Yadav}
\affiliation{%
 \institution{Indian Institute of Technology, Ropar}
 \country{India}}
\email{e.23csz0002@iitrpr.ac.in}
\author{Rahul Narava}
\affiliation{%
  \institution{Indian Institute of Technology, Ropar}
  \country{India}}
\email{syam.21csz0018@iitrpr.ac.in}
\author{Anshika}
\affiliation{%
  \institution{Indian Institute of Technology, Ropar}
  \country{India}}
\author{Shashi Shekher Jha}
\affiliation{%
  \institution{Indian Institute of Technology, Ropar}
  \country{India}}
\email{shashi@iitrpr.ac.in}

\renewcommand{\shortauthors}{Harshith et al.}

\begin{abstract}
Managing equal charge levels in active cell balancing while charging a Li-ion battery is challenging. An imbalance in charge levels affects the state of health of the battery, along with the concerns of thermal runaway and fire hazards. Traditional methods focus on safety assurance as a trade-off between safety and charging time. Others deal with battery-specific conditions to ensure safety, therefore losing on the generalization of the control strategies over various configurations of batteries. In this work, we propose a method to learn safe battery charging actions by using a safety-layer as an add-on over a Deep Reinforcement Learning (RL) agent. The safety layer perturbs the agent’s action to prevent the battery from encountering unsafe or dangerous states. Further, our Deep RL framework focuses on learning a generalized policy that can be effectively employed with varying configurations of batteries. Our experimental results demonstrate that the safety-layer based action perturbation incurs fewer safety violations by avoiding unsafe states along with learning a robust policy for several battery configurations.

\end{abstract}

\ccsdesc[500]{Computing methodologies~Sequential decision making}
\ccsdesc[500]{Computing methodologies~Markov decision processes}
\ccsdesc[300]{Applied computing~Electronics}


\keywords{Safe Reinforcement Learning, Battery Management System, Action Perturbation, SoC, Electric Vehicle}


\maketitle
 
\section{Introduction}
\begin{figure}[ht]
    \centering
    \includegraphics[scale=0.42]{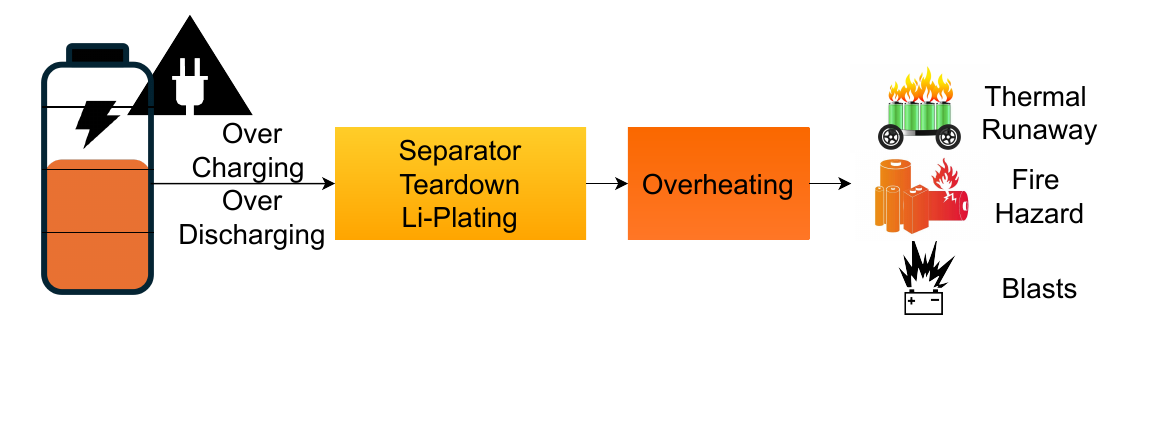}
    \caption{Safety Hazards in Li-ion Battery Charging}
    \label{fig:SRBMS}
\end{figure}

In recent times, the success of deep neural network-based Reinforcement Learning (RL) frameworks has shown their immense potential in several real-world applications~\cite{sallab2017deep}. Among those several applications, Deep RL methods have immense potential in Control Tasks. RL techniques have shown great results in the game domain~\cite{mnih2015human}, however, their application in real-world tasks becomes more tricky due to the fact that in the real world, several safety constraints need to be followed. One such application is the active cell balancing in Li-ion batteries.

Li-ion batteries are electrochemical energy storage systems with higher energy density and are easy to maintain. They're used in a wide range of devices, from mobile phones and laptops to electric vehicles (EVs). Electrochemical batteries, in general, are liable to get extremely hot, and Li-ion batteries, especially, are prone to explosion if the temperature conditions are beyond the safe threshold, as illustrated in Figure \ref{fig:SRBMS}. In extreme situations, the thermal runway can lead to irreversible battery damage and cause explosions, fires, and smoke. Li-ion batteries are the heart of EVs, and enhancing the battery life can make EVs a great asset for green and sustainable energy. Basically, the aim of balancing is to maintain a healthy state-of-charge (SoC) for every individual cell of the battery, thereby improving the battery's life and performance.

In Li-ion batteries, the cell balancing is done either by passive cell balancing or active cell balancing~\cite{paidi2022active}. While passive cell balancing is achieved by the use of resistors where extra charge is dumped from stronger cells in the form of heat, active cell balancing is more complex as the charge of stronger cells is shared by the weaker cells instead of wasting it into resistors. In this paper, we focus on a Deep RL-based strategy for active cell balancing. The charging time of a Li-ion battery depends significantly on the charging strategy employed. While making a charging strategy, several trade-offs need to be considered. One such trade-off is between the charging time and battery ageing. If a charging strategy uses current patterns that charge the battery aggressively by high input currents, there is a very high chance of battery degradation by several phenomena like lithium-plate deposition and Solid-Electrolyte Interphase (SEI) growth~\cite{tomaszewska2019lithium}. On the other hand, the adherence to safety parameters while charging a battery results in battery taking more time to charge. Using a battery in an EV with an aggressive charging strategy does not only require frequent replacements of the EV battery which is costly but also leads to higher chances of life-threatening incidents like thermal runaway and battery blasts~\cite{thermal2022ahmad}. The absence of a proper cell balancing technique makes the cells of a battery prone to getting unequal SoCs (and voltages). This poses many safety risks, ranging from decreased state-of-health (SoH) of cells to thermal runaway due to overcharging and deep-discharging in cells. Due to the risk it poses to the vehicle and human life, using a strategy that ensures safety becomes severely important. 

Traditional methods like constant current-constant voltage (CC-CV)~\cite{cope1999art} maintain safety constraints at the cost of higher charging time, which is undesirable. Hence, such a method does not provide an optimal charging policy, as the charging time is too long. RL techniques can be used to make the charging strategy safe and cost efficient by learning a safe and robust charging policy. 
For an RL setting, the battery pack acts as the environment wherein the learning agent takes actions from a continuous action space following a policy. The agent's action is the input current applied to the battery pack according to the system's constraints. The agent learns the policy by interacting with the battery pack environment making it capable of adjusting the policy accordingly as the battery ages. Also, the use of model-free RL \cite{park2022deep} allows obtaining a charging strategy without having a complete model of the battery using a feedback loop. However, enforcing safety constraints still remains essential for the RL-based strategy for practical use. It is crucial for the agent to learn a policy that ensures safety and does not violate constraints even during exploration. Learnig such a policy falls under the domain of safe exploration in RL that has become a prominent topic of research. In this paper, we attempt to learn a safe policy that performs active cell balancing and adheres to safety constraints of the battery. Our simulations show that the proposed deep RL-based safe-policy encounters less number of safety violations while converging faster to a balanced SoC in the battery.


The rest of the paper is organized as follows: In Section \ref{sec:lit_survey} we discuss about the traditional models and reinforcement learning models used for battery management systems previously. In Section \ref{sec:prelim} a walkthrough of the preliminaries is done where RL framework is briefly discussed. In Section \ref{sec:proposed}, we describe the work we proposed to achieve the objectives of the BMS through safe action perturbations. In Section \ref{sec:exp}, we discuss the implementation of the proposed approach. In Section \ref{sec:results} we present the evaluation results of our work. Finally, in Section \ref{sec:con} we conclude the paper.

\section{Related work}
\label{sec:lit_survey}
In this section, we discuss about the various works proposed for Battery Management Systems.
\subsection{Traditional Methods }
Several methods have been proposed  for an optimal charging policy but safety still remains an open problem. Several attempts have been made to solve this problem using Electrochemical Models (EMs), Mathematical models, and equivalent circuit models. Electrochemical Models (EMs) are model-based methods that rely on the electrochemistry of the battery, due to which it shows higher accuracy. 
Using an electrochemical model, Perez et al.~\cite{perez2015enhanced} presented a comprehensive study on the inverse relation between battery health and battery charging time and  a non-linear predictive control strategy for the same is used in \cite{zou2017electrochemical}. A quadratic dynamic matrix control model was given by Torchio et al. \cite{torchio2015real} as an attempt to give an optimal strategy, while \cite{pozzi2018film}presented a nonlinear predictive control strategy for checking Li-plate deposition and layer growth. Such model-based methods, though, show good accuracy but they come with several drawbacks. With the ageing of the battery, its parameters change, which presents another challenge in developing a strategy that can adapt to the changes due to ageing. Also, EMs consist of a large number of states, most of which are not even measurable practically, which makes it challenging to access the full information about the state. Moreover, due to the large number of states in the EMs,it turns into a large-scale optimisation problem. 

The alternative approach to tackling these limitations of model-based methods is to use model-free methods. Several approaches have been proposed, the most prominent of which are the rule-based CC-CV approach \cite{cope1999art} and the use of the Kalman filter for battery SoH estimation. Another approach is a Machine Learning model for fast battery charging techniques given by  Attia et al. \cite {attia2020closed}, where battery life is maximized by current profile parameterization and further use of Bayesian optimization to find the optimal charging sequence. In order to tune the CC-CV method to take temperature constraints into account, Patnaik et al. \cite{patnaik2018closed} proposed constant current, constant temperature, and constant voltage (CC-CT-CV), which is basically a closed-loop charging strategy. However, like model-based methods, there are several drawbacks to model-free methods as well. The main issues include accuracy, the lack of a guarantee that the charging policy obtained will be optimal, adaptation of the policy to the change in battery parameters, and the need for hit-and-trial to obtain an optimal policy.
\subsection{RL Based Methods}


The authors in \cite{park2022deep} address the fast charging problem of Li-ion batteries using DDPG as their Deep RL method. \cite{ppobat} has addressed the problem of balancing the SoC and cell temperature of Li-ion batteries. However, the drawback of both approaches \cite{park2022deep}, \cite{ppobat} is that they don't account for the safety violations. To incorporate safety in the context of RL, a well-known problem is safe exploration. Safe exploration refers to the exploration in which the system avoids or never enters an unsafe state. Given a deterministic environment, basic RL methods can ensure safety to a great extent. However, in a non-deterministic system like ours, traditional methods don't work too well.
Hans et al. \cite{hans2008safe} identified two other components of safe exploration, namely, \textit{safety function} and \textit{backup policy}. The function that defines how safe an action is in a given state is regarded as the \textit{safety function}. Pre-determining this regarding an action can help take an action that doesn't lead to an unsafe transition to a dangerous state. \textit{Backup policy,} on the other hand, is a policy that can give a safe action and can bring the agent to a safe region. It is important to mention that a safe policy may not necessarily be an optimal one and vice-versa. Dalal et al. \cite{dalal2018safe} presents a safety-layer method that analytically makes action corrections based on each state to ensure that the RL learned policy doesn't violate constraints in general physical environments. 

In this paper, from the prior literature, we employ a safety layer that perturbs the action before acting upon the environment to ensure safety while exploring in the battery setting.
\section{Preliminaries}
\label{sec:prelim}
This section is divided into three parts, where Section \ref{sec:3.1} discusses the framework of RL and its adaptation to our battery environment and Section \ref{sec:3.2} discusses the Battery Electrochemical-Thermal model, and Section \ref{sec:3.3} refers to the workflow of DDPG.
\subsection{Basics of Reinforcement Learning (RL)}
\label{sec:3.1}
In this work, we formulate the problem using Markov Decision Process (MDP), a mathematical decision-making tool that can be applied to solve complex RL problems.
A MDP\cite{Sutton1998} is a tuple of $ < S, A, P, R, \gamma>$ where we define the environment, the state space $S$, action space $A$ (which may be discrete or continuous), transition dynamics $P : S \times A \rightarrow S$, reward function $R$ and the discount factor $\gamma \in [0, 1]$. Apart from normal MDP, A constrained MDP is a tuple of $ < S, A, R, P, C, \gamma>$ that also includes a constraint function $c$ which we use to involve safety aspect in the RL Framework  

In our setting, we use a constrained MDP\cite{altman2021constrained} tuple $ < S, A, R, P, C, \gamma>$ where $S$ is the continuous state space of the battery environment, $A$ is the continuous action space, $P$ is transition probability, which is accessible only if there is full access to the model of the environment, which is not the case in our environment, whereas $C$ is the constraint function.


A policy $\pi: S \rightarrow A$ determines what action to take give a state at each time step. The objective of any RL problem is to learn an optimal policy. Further, a run of the policy $\pi$ generates a trajectory which is a sequence of states and actions $\tau = (s_0, a_0, r_0, s_1, a_1, r_1, \ldots)$ where $s_{t+1} \sim P(\cdot\mid s_t, a_t)$ and $r_t = r(s_t, a_t)$.
Basically, the agent performs an action \(\mathbf{a_t} \in \mathbf{A}\) on the environment \(\mathbf{E}\), i.e., the Battery environment based on a policy. The configuration of the environment, at every step, is defined by the state \(\mathbf{s_t} \in \mathbf{S}\). After the action is performed on the environment, its configuration transitions to a new state \(\mathbf{s\textsubscript{t+1}} \) and the environment generates a reward \(\mathbf{r_t} \in \mathbb{R}\). This reward depends on the current state and current action and is used to define $\mathbf{R}$ which is the total discounted return, calculated as: 
\begin{equation}
R_t = \sum_{k=0}^{\infty} \gamma^k r(s_{t+k}, a_{t+k})
\end{equation}

In order to get the optimal policy $\pi^{*}$ for a model-based setting, the maximum of the value function is taken as: 
\begin{equation}
    \pi^* = \arg \max_{\pi} V^{\pi}(s_t)
\end{equation}

Here, $V^{\pi}(s_{t})$, known as the value function, is the expected total discounted return
\begin{equation}
    V^{\pi}(s_t) = \mathbb{E}[R_t \mid s_t]
\end{equation}

In model-free settings like the Battery environment in this paper, the state-action value function should be maximized. It is defined as:
\begin{equation}
    Q^{\pi}(s_t) = \mathbb{E}[R_t \mid s_t, a_t]
\end{equation}
 i.e., the expectation of total discounted reward following a policy $\pi$, given that action $a_{t}$ is taken in the state $s_{t}$. Hence, the following equation holds: 
 \begin{equation}
V^*(s_t) = \max_{a_t \in A} Q^*(s_t, a_t)
\end{equation}
In every state, to know which action to take, the \(Q\) value is maximized over action space \(\mathit{A}\) in the following manner:
\begin{equation}
a^*_t = \arg \max_{a_t \in A} Q^*(s_t, a_t)
\end{equation}

\begin{figure*}[htbp]
    \centering
    \includegraphics[scale=0.4]{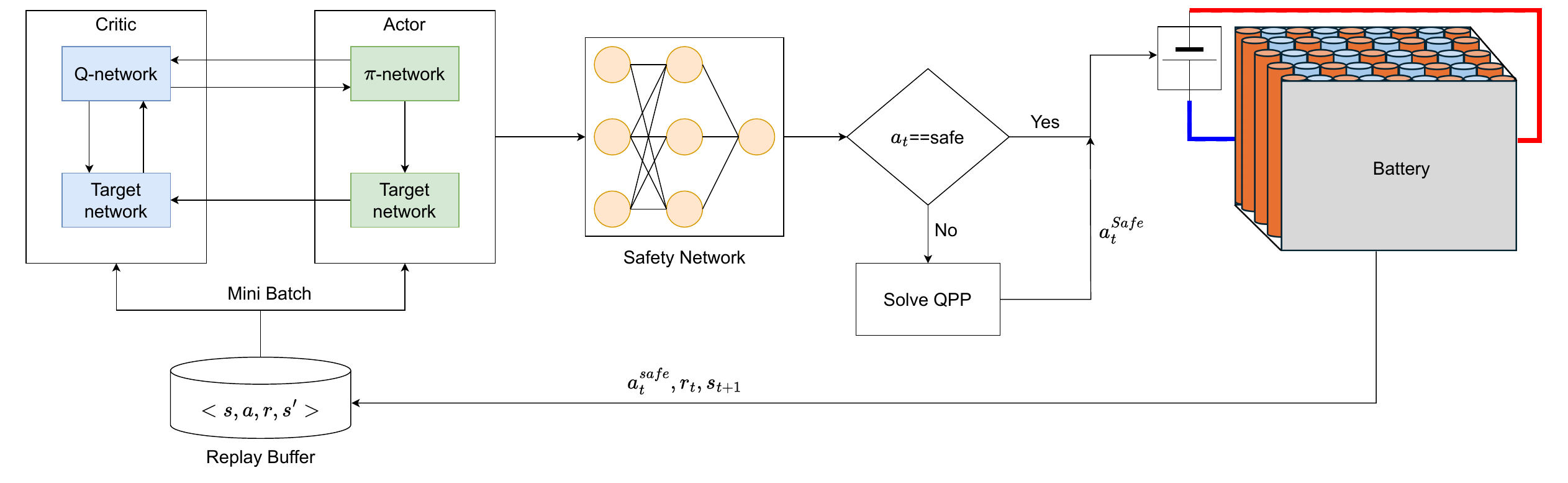}
    \caption{The flow diagram of the proposed approach}
    \label{workflow}
\end{figure*}
\subsection{Battery Electrochemical-Thermal Model}
\label{sec:3.2}
The Battery environment is very complex, with several metrics involved at chemical and electrical levels. Park et al.\cite{park2022deep} attempted to present a computational model of a Li-ion battery that uses the Doyle–Fuller–Newman (DFN) model. It is a thorough framework for simulating several physical events inside a battery. The Porous electrode theory \cite{smith2017multiphase} serves as the foundation of this model. The phenomena that get simulated using the DFN model, are Lithium Concentration in Solid Phase \( c_s^\pm (x, r, t) \), Lithium Concentration in Electrolyte \( c_e (x, t) \), Battery Temperature \(T(t)\), Molar Influxes \(j_n^\pm (x, t)
\), Ionic Current \(i_e^\pm (x, t)\),  Solid Electric Potential \(\phi_s^\pm (x, t)\), Electrolyte Electric Potential \(\phi_e (x, t)\).
On a high level, the governing equation for Battery Temperature \(T(t)\) modeling is: 
\begin{equation}
mc_{P} \frac{dT}{dt} (t) = \frac{1}{R_{\text{th}}} [T_{\text{amb}} - T(t)] + \dot{Q}
\end{equation}
and for the Heat Generation rate \(\dot{Q}\) is:
\begin{equation*}
     \dot{Q} = I(t) \left[ U^+ (t) - U^- (t) - V(t) \right] - I(t) T(t) \frac{\partial}{\partial T} \left[ U^+ (t) - U^- (t) \right]
    \label{HG}
\end{equation*}

Here, \(m\) is the mass, \(c_P\) is the specific heat capacity, \(R_{\text{th}}\) is the thermal resistance, \(T_{\text{amb}}\) is the ambient temperature, \(I(t)\) is the applied current, and \(U^\pm (t)\) are the open-circuit potentials of the electrodes.

In the battery, the Li-ion concentration, Temperature of the battery, Voltage of the battery and the State of Charge (SoC) broadly depend on Li-ion concentration in a manner that takes into consideration the average Li-ion concentration on the positive and negative electrodes, along with the electrolyte concentration. Temperature is a critical metric for the safety and performance of the battery and voltage is an essential measure for battery SoC and SoH. SoC is a metric that acts as the fuel gauge in a battery scenario. It indicates the relative charge left in the battery compared to the total charge. Factors like temperature, ageing and discharge rates affect the SoC. Among several methods to measure SoC, the simplest method is Coulomb-Counting in which the amount of charge entering and leaving the battery is directly measured by integrating the current over time.

\subsection{DDPG}
\label{sec:3.3}
The RL-based agent employed in our proposed approach is the Deep Deterministic Policy Gradient (DDPG) \cite{lillicrap2015continuous} method. DDPG is well-suited for environments that has continuous state and action spaces. As there are infinite actions possible, it is not possible to compute learned probabilities for each of the actions, hence statistics of probability distribution like mean and variance are learned. In DDPG, the policy is parameterized as long as $\nabla\pi(a \mid s, \theta)$ exists. Further, an action $\mathbf{a}$ in state $\mathbf{s}$ is given as $a = \mu_\theta(s)$~\cite{silver2014deterministic}. The deterministic policy gradient to train the DDPG network is given as: 
\begin{equation}
\nabla_\theta J(\mu_\theta) = \mathbb{E}_{s \sim \rho_{\mu}} \left[ \nabla_\theta \mu_\theta(s) \nabla_a Q^{\mu}(s, a) \bigg|_{a = \mu_\theta(s)} \right]
\end{equation}
where $\mu_{\theta} : \mathbf{S\to A } $, $ \theta \in \mathbb{R}^n $ is the parameter to be learnt, $J(\mu_\theta) $ is the expected cumulative discounted sum of rewards starting from the initial state. The policy gradient is valid given $ \nabla_\theta \mu_\theta(s) $, $\nabla_a Q(s, a)$ exist. \\

In DDPG, an actor-critic approach is applied that have two networks an Actor deep neural network and a Critic deep neural network. In order to enforce safety, the safety constraints are used as soft constraints with penalties received on bad transitions. As the DDPG agent learns a deterministic policy, corresponding to each state of the battery pack, the agent generates an action. Both actor and critic have their respective target networks that make learning stable. 
With the on-policy exploration for deterministic policy, it may be possible that only a few actions are explored. For thorough exploration, a type of Noise may be Gaussian or Ornstein–Uhlenbeck \cite{uhlenbeck1930theory} is used. 
Further, for a wide range of experiences, stable behaviour and avoiding over-fitting, Experience Replay Buffers are used that store the previous experiences and a mini-batch can be sampled randomly for updates.

\section{The Proposed Approach}
\label{sec:proposed}
In this section, we discuss our proposed deep RL based method that adheres to safety constraints while ensuring fast charging during the charging of Li-ion batteries.

As shown in Figure \ref{workflow}, the agent perceives a state $s_t$and takes an action based on safety network $\mathcal{S}_{\theta^\mathcal{S}}$ and policy $\mu_{\theta^{\mu}}$. To ensure safety, the action from the policy is evaluated based on the safety network. If it's an unsafe action, then the action perturbation is done by solving a Quadratic Programming Problem (QPP). The agent transitions to the next state $s_{t+1}$ and receives a reward $r_t$ using $a_{t}^{safe}$. These samples are stored in a replay buffer $\mathcal{R}$. We sample a minibatch from the buffer $\mathcal{R}$ to update the actor-critic networks to learn a safe policy $\mu_{\theta^{\mu}}$.
\subsection{MDP Formulation}
In our setting, the Markov Decision Process is formulated in the following manner:\\
\textbf{States}: The state of the battery environment is of size 3 and is defined using three metrics, the \textit{temperature, voltage, SoC}. Since, in this model, we don't deal with a specially configured battery, so we don't have other internal battery parameters for learning the policy. The training is done using the output-based metrics mentioned before.\\
\textbf{Action}: The action for battery charging is the input current, which is a scalar. Further, at every time step, the action given by DDPG is evaluated by the safety layer, and a safe action is given in \textit{Amperes}.\\
\textbf{Rewards}
The reward that we use is the sum of the reward for violating constraints and the reward for increasing time steps.
\begin{equation}
r_t = r_{violation} + r_{timestep}
\end{equation} 
where   $  r_{violation} = -100 * V_{violation} - 5 * T_{violation}$
The reward for the time step has been set to -0.1 for every step. This reward ensures that the learned policy ensures fast charging as well.
\subsection{Safety Action Perturbations}
The DDPG algorithm, as discussed in Section \ref{sec:3.3}, is employed on the MDP presented above. This outputs a single action as the input current. The input current determines the voltage, temperature, and SoC of the battery in a particular state. We make sure that the action employed on the battery is a safe action i.e. it does not lead to a state where the voltage and temperature of the battery are beyond the maximum safe limits. For this, we introduce a safety network that generates a safety signal \cite{dalal2018safe}. 

Suppose the DDPG algorithm outputs the action $a_t$ at any time step $t$. Before executing the action $a_t$ in the environment, the safety signal $G$ is generated from the safety network based on the current state conditions $s_t$. This safety signal $G$ has same dimensions as of the action. A Quadratic Programming Problem (QPP) is solved using this safety signal to perturb the original action $a_t$ that outputs a safe action $a_{safe}$. The QPP is formulated as follows: 
\begin{equation}
\begin{aligned}
& \text{minimize} & \frac{1}{2} \mathbf{a}_{\text{safe}}^T \mathbf{H} \mathbf{a}_{\text{safe}} + \mathbf{a}^T \mathbf{a}_{\text{safe}} \\
& \text{subject to} & \mathbf{C}\mathbf{a}_{\text{safe}} \leq \mathbf{d}
\end{aligned}
\label{qpp}
\end{equation}\\
Here \(\mathbf{H}\) is the identity matrix and \(\mathbf{C}\) is the constraints matrix (transpose of \(\mathbf{G}\)). \(\mathbf{d}\) represents the threshold for the safety limits. 

In our case, we make sure that the action (input current) does not go beyond the threshold of $-4.2 \space Ampere$ \cite{park2022deep}, where the negative sign only denotes the reversed direction of the current. 
This approach guarantees a valid action because of the properties of the QPP itself and the nature of optimization.

The objective function in equation \ref{qpp} is a convex function, which ensures that the local and global minimum are the same, making it easier to find a feasible solution. Also the constraints to which the objective function is subjected to,  $\mathbf{C}_{\mathbf{a}_{\text{safe}}} \leq \mathbf{d}$ form a convex feasible region. All of this ensures that a feasible solution exists, and we get an action \(\mathbf{a_{safe}}\) that adheres to the safety constraints.

Finally, the safe action obtained as a result of solving this QPP is executed in the environment. The consequent experience is then stored in the replay buffer as $(s_t, a_t^\text{safe}, r_t, s_{t+1})$.
The Safety Network $\mathcal{S}$ is then updated with a loss function given as:
 \begin{equation}
      L_S = \frac{1}{N} \sum_i (\mathcal{S}(s_i|\theta^\mathcal{S}) - \text{K}_i(s))^2
      \label{eq: two}
 \end{equation}
 Let $T(s)$ be the temperature component and $V(s)$ be the voltage component of a state $s$. Then the safety targets $K_i(s)$ are calculated as: 
 \begin{equation}
    K_i(s) = 
\begin{cases} 
0 & \text{if } T(s) < 35 \text{ and } V(s) < 4.2 \\
1 & \text{if } T(s) \geq 35 \text{ or } V(s) \geq 4.2 
\end{cases}
\label{targetcal}
 \end{equation}
Finally, the parameters are updated using a gradient descent step as follows: 
 \begin{equation}
     \theta^\mathcal{S} \leftarrow \theta^\mathcal{S} - \alpha_S \nabla_{\theta^\mathcal{S}} L_S
     \label{three}
 \end{equation}
           
The proposed approach has been depicted in the Algorithm \ref{alg:alg}. We initialize the actor-critic networks, replay buffer $\mathcal{R}$, and the safety network $\mathcal{S}$. To ensure safety while exploration, the actions are perturbed based on the safety network by solving a QPP. Then the samples are stored in the replay buffer $\mathcal{R}$ by running the policy in the environment. Finally, A mini-batch of samples is retrieved from the replay buffer $\mathcal{R}$ to update the actor-critic networks in order to learn a safe policy.
\begin{algorithm}
\caption{Safe Action Perturbation for Active Cell Balancing} \label{alg:alg}
\begin{algorithmic}[1]
    \State Initialize Replay Buffer $\mathcal{R}$, Actor $\mu(s|\theta^\mu)$, Critic $Q(s, a|\theta^Q)$
    \State Initialize Target Networks $\mu'$ and $Q'$ with $\theta^{\mu'} \leftarrow \theta^\mu$, $\theta^{Q'} \leftarrow \theta^Q$
    \State Initialize Safety Network $\mathcal{S}(s, a|\theta^\mathcal{S})$
    \For{episode $= 1$ to $M$}
        \State Reset environment, get initial state $s_0$
        \For{step $= 1$ to $T$}
            \State Select action $a_t = \mu(s_t|\theta^\mu) + \mathcal{N}_t$
            \State Evaluate Safety Signal $G_t = \mathcal{S}(s_t, a_t|\theta^\mathcal{S})$ using \ref{qpp}
            \State Execute $a_t^\text{safe}$, observe $r_t$ and $s_{t+1}$
            \State Store $(s_t, a_t^\text{safe}, r_t, s_{t+1})$ in $R$
            \State Sample mini-batch $(s_i, a_i, r_i, s_{i+1})$ from $R$
            \State Set target for Critic:
            \[
            y_i = r_i + \gamma Q'(s_{i+1}, \mu'(s_{i+1}|\theta^{\mu'})|\theta^{Q'})
            \]
            \State Update Critic using target:
            \[
            L = \frac{1}{N} \sum_i (y_i - Q(s_i, a_i|\theta^Q))^2
            \]
            \State Update Actor using sampled policy gradient:
            \[
            \nabla_{\theta^\mu} J \approx \frac{1}{N} \sum_i \nabla_a Q(s, a|\theta^Q) \nabla_{\theta^\mu} \mu(s|\theta^\mu)
            \]
            \State Compute safety targets $\text{K}_i$ using \ref{targetcal}

            \State Update Target Networks:
            \[
            \theta^{Q'} \leftarrow \tau \theta^Q + (1 - \tau) \theta^{Q'}
            \]
            \[
            \theta^{\mu'} \leftarrow \tau \theta^\mu + (1 - \tau) \theta^{\mu'}
            \]
            \State Update Safety network by minimising $L_S$ using \ref{three}
            
        \EndFor
    \EndFor
\end{algorithmic}
\end{algorithm}

\section{Experimentation}
\label{sec:exp}
An overview of the simulation environment, deep network architectures, and the configurations used to evaluate our proposed approach against the baseline method is given in this section. 
\subsection{Training the battery SoC}
The battery environment is a simulated counterpart of the electrochemical thermal model of a Li-ion Battery which has already been validated against a real battery pack by Park et al. \cite{park2022deep}. This battery environment simulates phenomena like electrode behaviour, electrochemical reactions along with heat generation. With the basic battery parameters intact, we utilized three different configurations for training. These configurations varied in the thickness of electrodes, the thickness of the separator, maximum voltage, minimum voltage, initial temperature, initial voltage, radii of particles at the electrodes, and specific inter-facial surface area. 

The goal of the agent is to learn a policy that converges faster to a balanced SoC in the battery while adhering to the safety constraints.
The policy is trained on 3 different configurations such that at every episode, the battery configuration is chosen uniformly among 3 configurations, and the results are averaged over 5 seed values. This ensures that the policy learned is robust and can be applied to several battery configurations.

\subsection{Deep Network Architectures}
We define the deep neural network architecture used for the DDPG and the Safety Layer Network. For DDPG, both the actor and critic consist of two hidden layers. The first hidden layer consists of 400 nodes, and the second hidden layer consists of 300 nodes in both networks. For mapping states to actions, the input to the actor-network is a state of size 3, and the output is a single action. 

The safety network, which is used to approximate the safety signal based on the current state of the environment, consists of two fully connected hidden layers. The first hidden layer consists of 128 nodes and takes the input as the state vector. The second hidden layer also contains 128 nodes, and the final fully connected output layer gives a tensor representing the safety signal. Both layers use ReLU activation to process their respective inputs. The hyperparameters used to train the agent are given in Table~\ref{table:hyperparm}.
\begin{table}[H]
\begin{tabular}{|l|r|}
\hline
\textbf{Hyperparameter}        & \textbf{Value} \\ \hline
Discount Factor $\gamma$       & 0.99           \\ \hline
Replay Buffer Size                 & 100000            \\ \hline
Minibatch size             & 64        \\ \hline
Critic Learning Rate            & 0.001          \\ \hline
Safety Learning Rate           & 0.001          \\ \hline
Actor Learning Rate           & 0.0001          \\ \hline
Target Update Parameter $\tau$ & 0.001          \\ \hline
\end{tabular}
\vspace{2mm}
\caption{Hyperparameters used for training.}
\label{table:hyperparm}
\end{table}

\subsection{Performance Metrics}
We compare the proposed approach against the baseline park et al.~\cite{park2022deep} using the following four metrics during the training: temperature violations, voltage violations, cumulative returns, and charging times per episode. These metrics are then averaged over five seed values for a thorough comparison for 3000 episodes. 

If the temperature is above the safety threshold level, the temperature violations record the difference between the current temperature and 
safe temperature as 
\begin{equation}
    \mathbf{T_{violation} = T_{curr} - T_{safe}}
    \label{eqntempviol}
\end{equation}
Similarly, the voltage violation captures the differences between the voltage and the maximum safe voltage as 
\begin{equation}
    \mathbf{V_{violation} = V_{curr} - V_{safe}}
    \label{eqnvoltvio}
\end{equation}

To analyze the safety violations of the policies learned by our approach with the baseline park et al.\cite{park2022deep}, we plot the number of safety violations per episode. 
For the number of violations per episode, the moving average with a window size of 10 episodes is plotted.
The plots are further analyzed to compare the effect of the safety layer on the learned policy. 

\section{Results}
\label{sec:results}
We analyze various metrics of the policy while training and also evaluate the policy on different conditions in this section,

\subsection{Performance during Training}
\begin{figure}[t]
\centering

\begin{minipage}[b]{.48\linewidth}
  \centering
   \includegraphics[width=\textwidth]{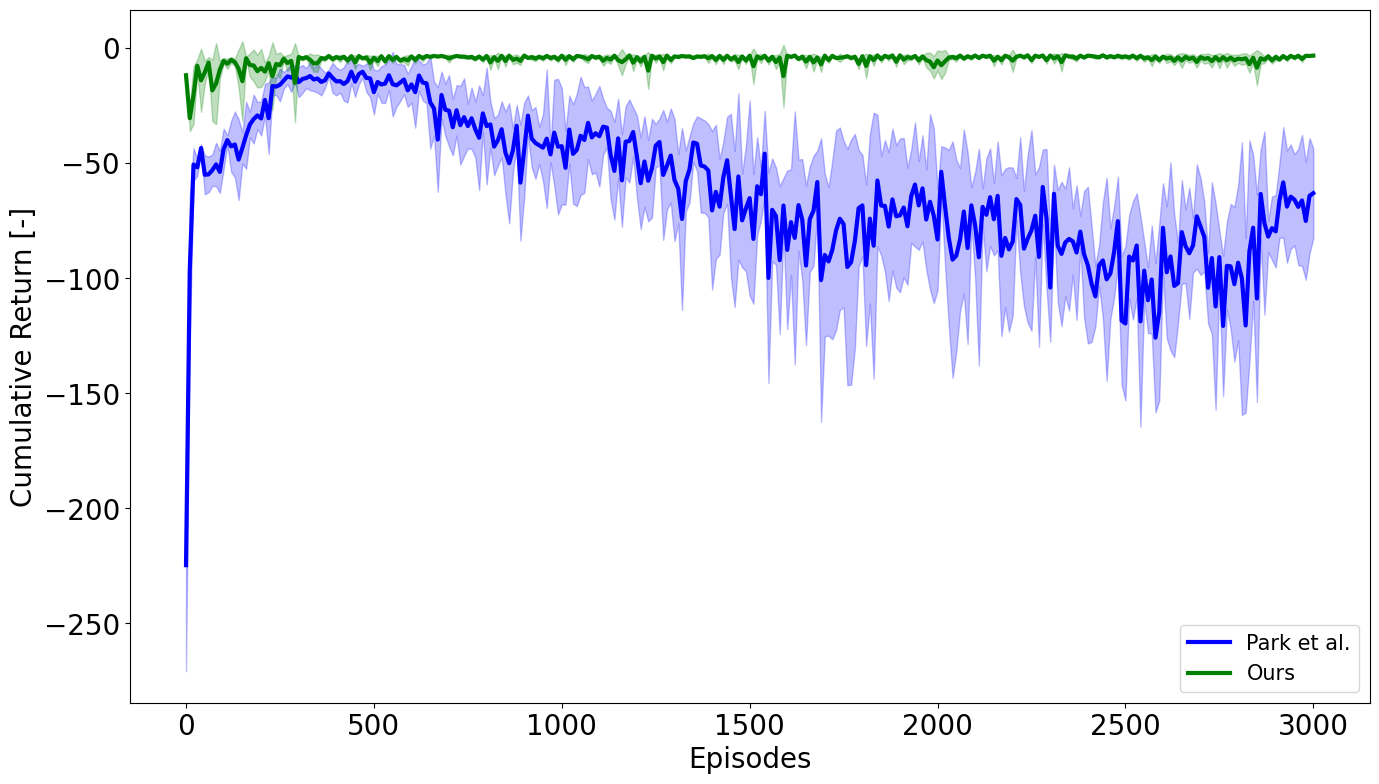}
    \label{fig:CR}
  \centerline{(a)}\medskip
\end{minipage}
\hfill
\begin{minipage}[b]{0.48\linewidth}
  \centering
  \includegraphics[width=\textwidth]{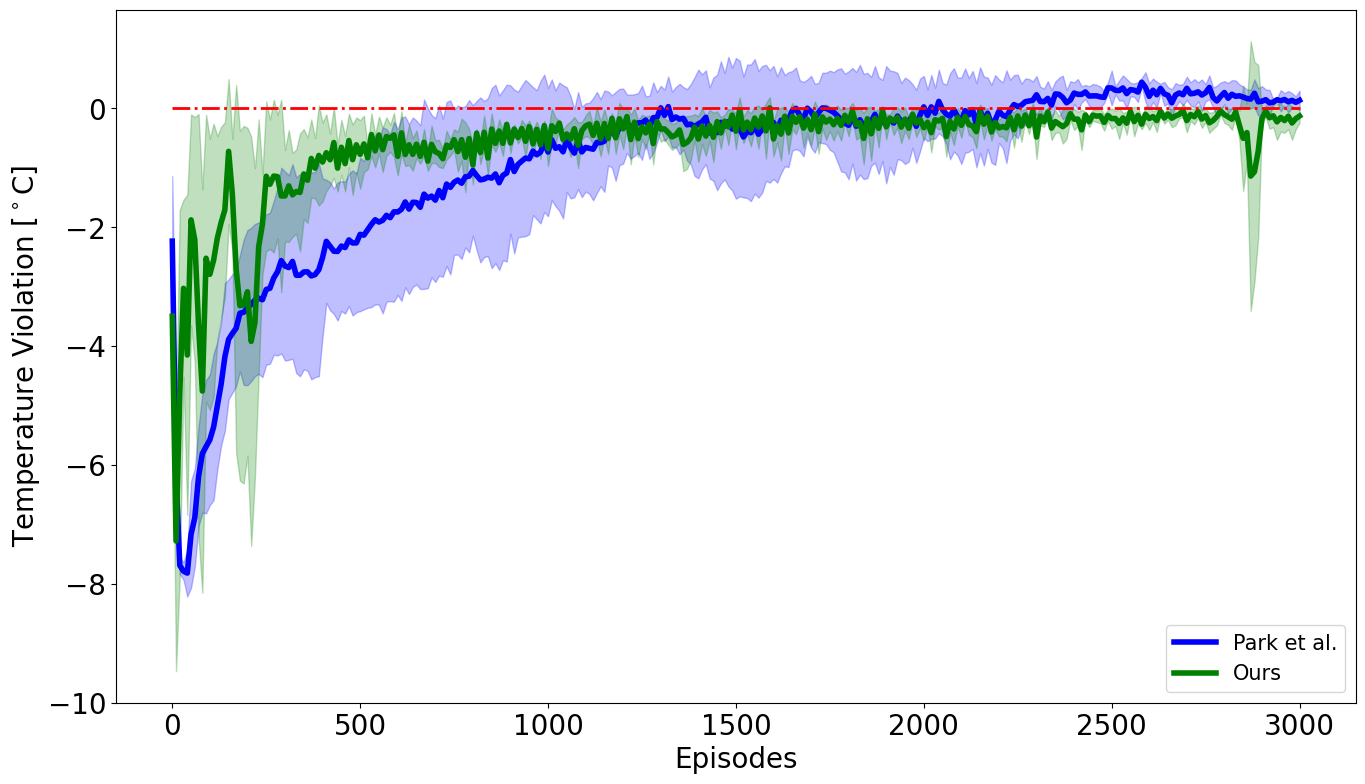}
    \label{fig:TV}
  \centerline{(b)}\medskip
\end{minipage}
\begin{minipage}[b]{.48\linewidth}
  \centering
  \includegraphics[width=\textwidth]{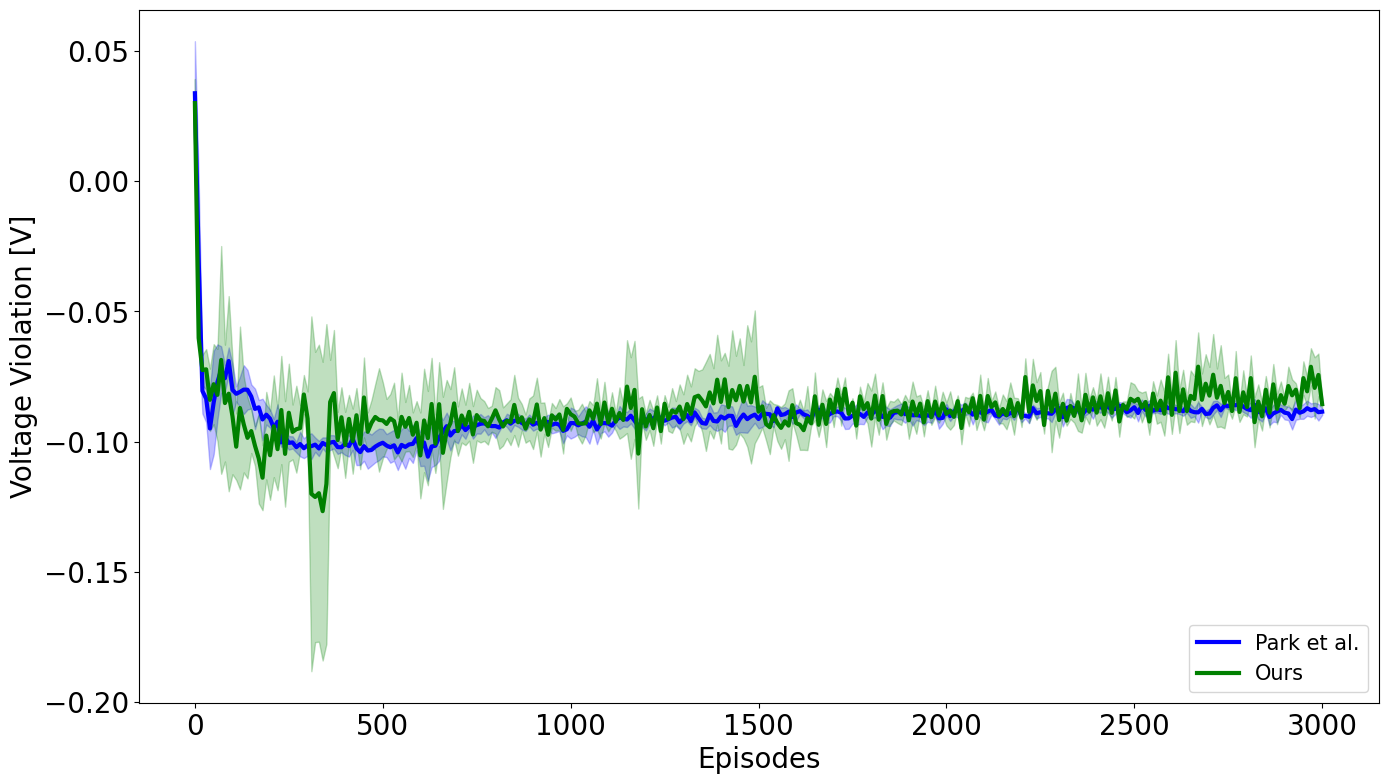}
    \label{fig:VV}
  \centerline{(c)}\medskip
\end{minipage}
\hfill
\begin{minipage}[b]{0.48\linewidth}
  \centering
  \includegraphics[width=\textwidth]{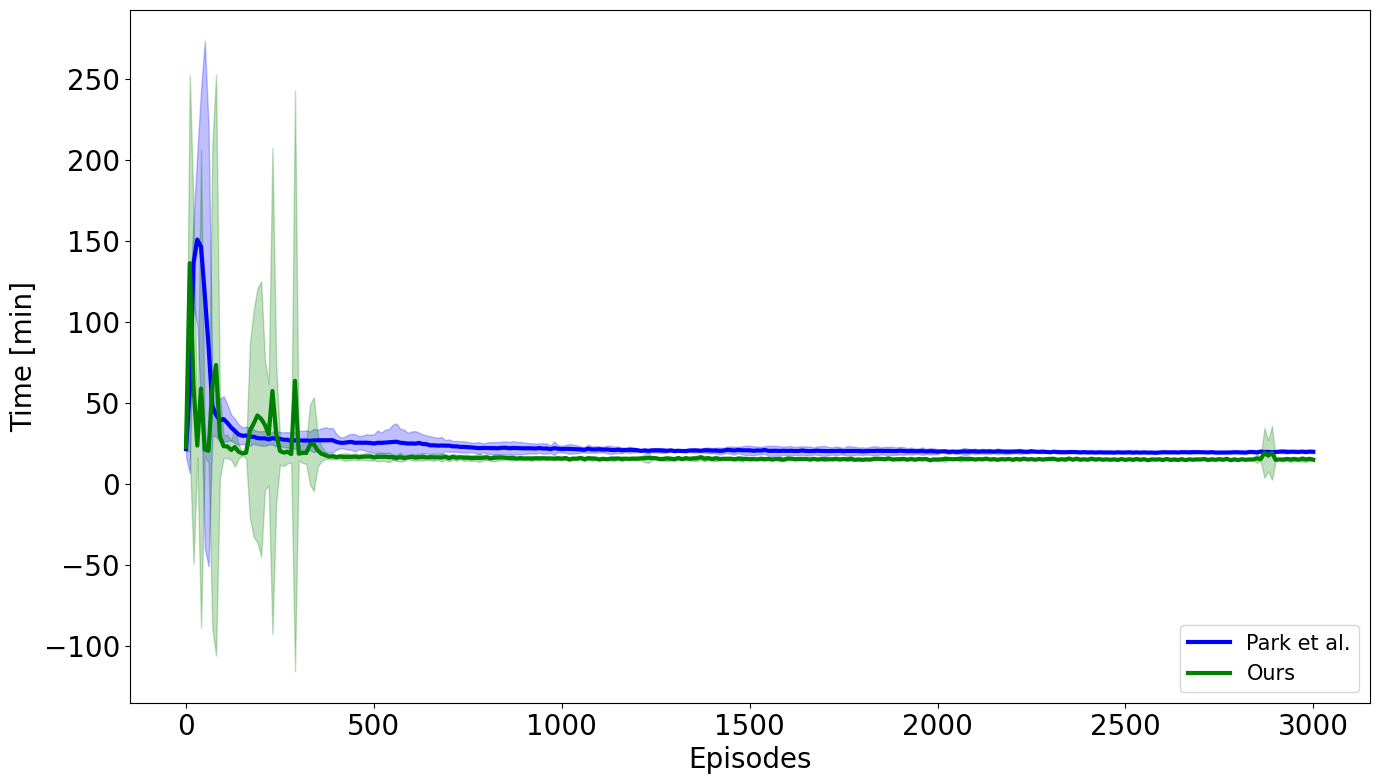}
\label{fig:CT}
  \centerline{(d)}\medskip
\end{minipage}
\caption{Performance metrics during training}
\label{allmetrics}
\end{figure}

\begin{figure}[H]
  \begin{minipage}[htbp]{0.45\textwidth}
        \centering
        \includegraphics[width=\textwidth]{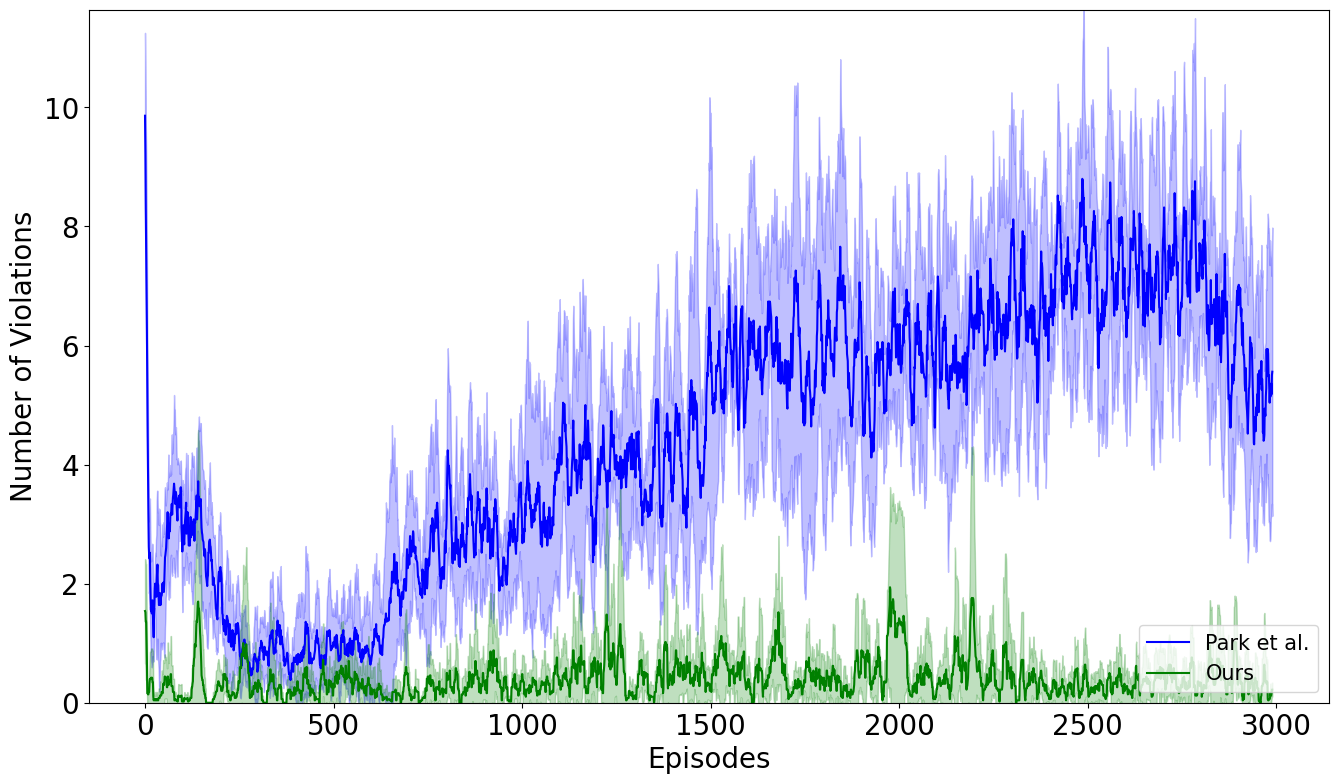}
        \caption{Total number of violations averaged over 5 seeds}
        \label{fig:NUMV}
    \end{minipage}
\end{figure}

The plots in Figures \ref{allmetrics}(a-d) and Figure \ref{fig:NUMV} depict various metrics on which the training performance has been evaluated.

When compared to the baseline that uses conventional DDPG, as shown in Figure \ref{allmetrics}(a), the cumulative return for the suggested DDPG with Safety Layer converges substantially earlier. Furthermore, the average cumulative return for the proposed DDPG with Safety Layer over the course of five seeds is \textbf{-4.93}, but the return is \textbf{-59.4} in the absence of it.
The temperature violations plot in Figure \ref{allmetrics}(b) shows that the proposed approach never crosses 0, i.e. the temperature does not exceed the maximum safe temperature constraint, in contrast to the policy with only DDPG. Therefore, we can say that there is no temperature violation for the proposed DDPG with the Safety layer.
Figure \ref{allmetrics}(c) illustrates the instances in which the voltage above the maximum safe voltage during training, akin to temperature violations. 
The charging time used by the policy to charge the battery to 80\% of the SoC level is the fourth performance evaluation metric. The charging time converges to a lower value even when safety requirements are met, as seen in Figure \ref{allmetrics}(d). The mean charging time for the standard DDPG is \textbf{23.89 minutes}, averaged over all seed values, whereas our method's mean charging time is \textbf{17.86 minutes}.

Finally, we also plot the number of constraint violations that take place during the training in both cases. This Figure \ref{fig:NUMV} clearly shows that simple DDPG violates a lot of safety constraints than the proposed method. On average, over all the seed values, for 3000 episodes, DDPG faces around \textbf{4.45 safety violations} per episode, while our proposed method only \textbf{0.394 safety violations} per episode.

\subsection{Performance of the learnt Policy during Evaluation}
\begin{figure}[htbp]

    \begin{minipage}[b]{0.45\textwidth}
        \centering
        \includegraphics[width=\textwidth]{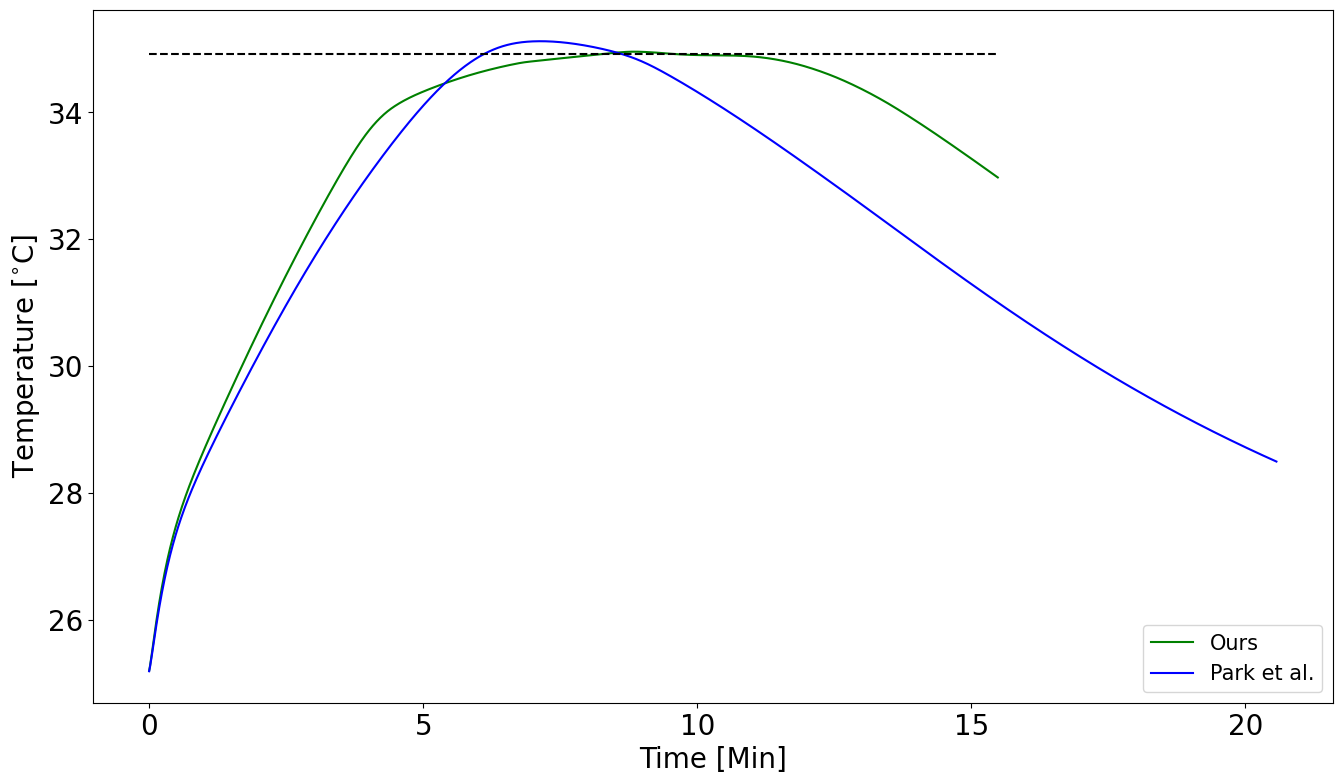}
        \label{fig:TEMP1}
        \centerline{(a)}\medskip
    \end{minipage}
 
    \begin{minipage}[b]{0.45\textwidth}
        \centering
        \includegraphics[width=\textwidth]{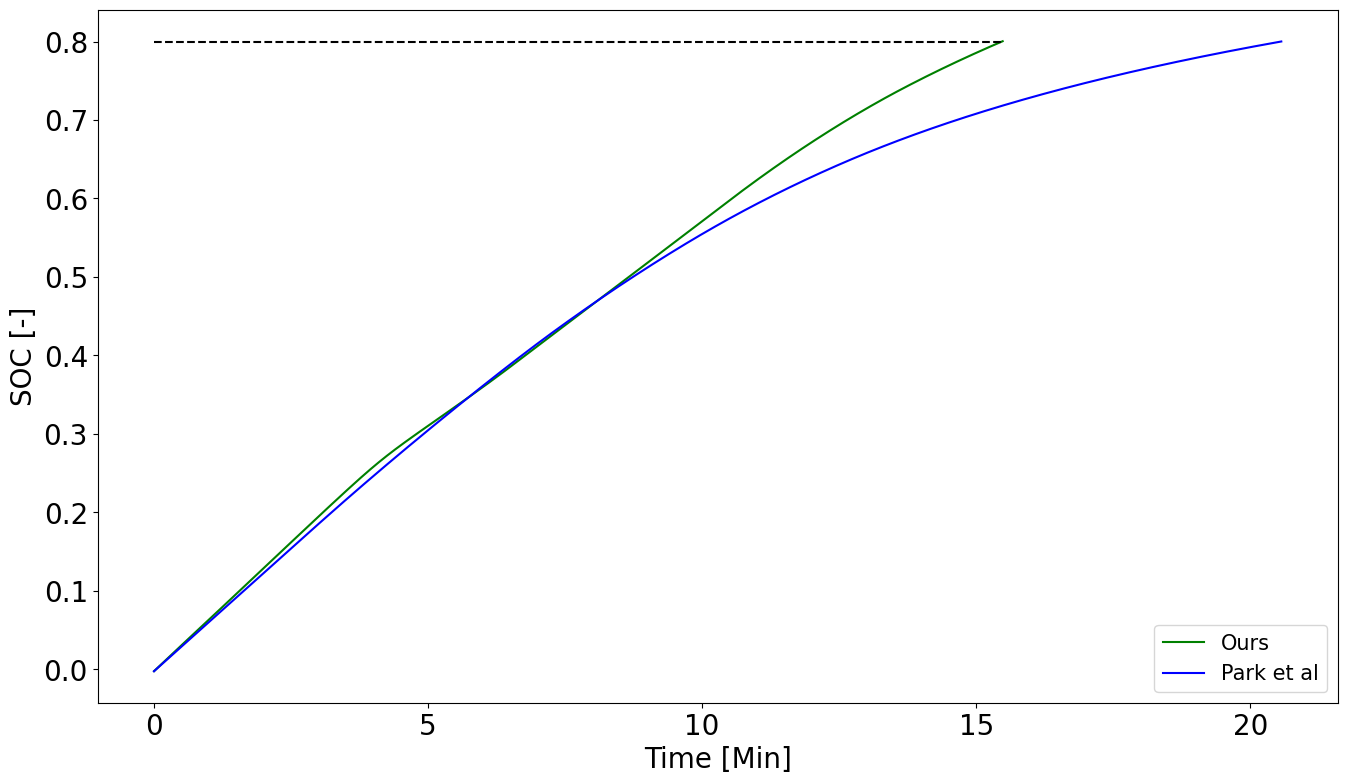}
        \label{fig:SoC1}
        \centerline{(b)}\medskip
    \end{minipage}
    \begin{minipage}[b]{0.45\textwidth}
        \centering
        \includegraphics[width=\textwidth]{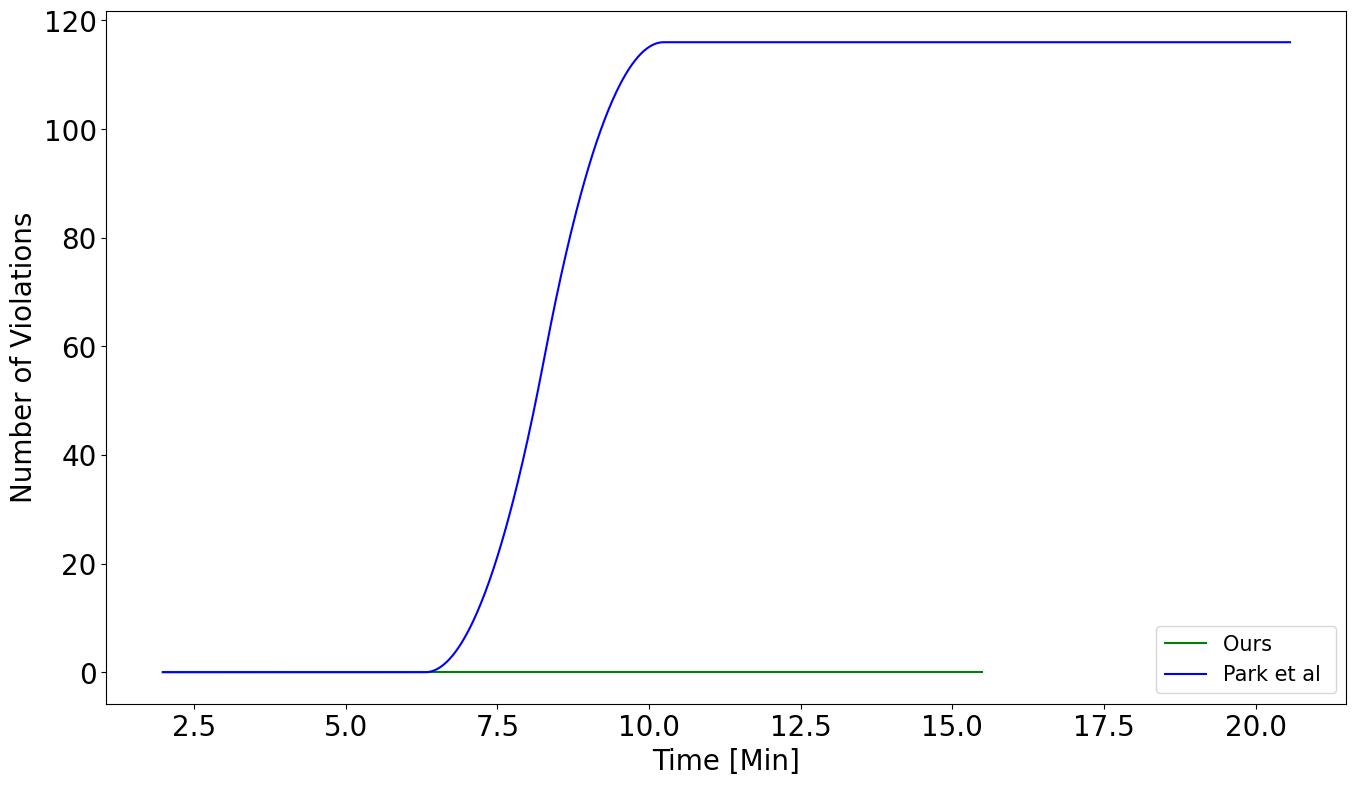}
        \label{fig:VIOALL1}
        \centerline{(c)}\medskip
    \end{minipage}
    \caption{Evaluation of proposed policy on configuration 1 with Initial Temperature=273K and Initial Voltage=2.5V (a)Temperature (b)SoC (c)Violations}
    \label{fig:config 1}
\end{figure}

In this section, we have assessed our proposed approach with the baseline in two distinct configurations varied in battery characteristics, noise levels, and starting circumstances.
The curves for the first configuration, with an initial voltage of 2.5 volts and an initial battery environment temperature of 273 Kelvin, are plotted in Figure \ref{fig:config 1}. 

\begin{figure}[htbp]
    
    \begin{minipage}[t]{0.45\textwidth}
        \centering
        \includegraphics[width=\textwidth]{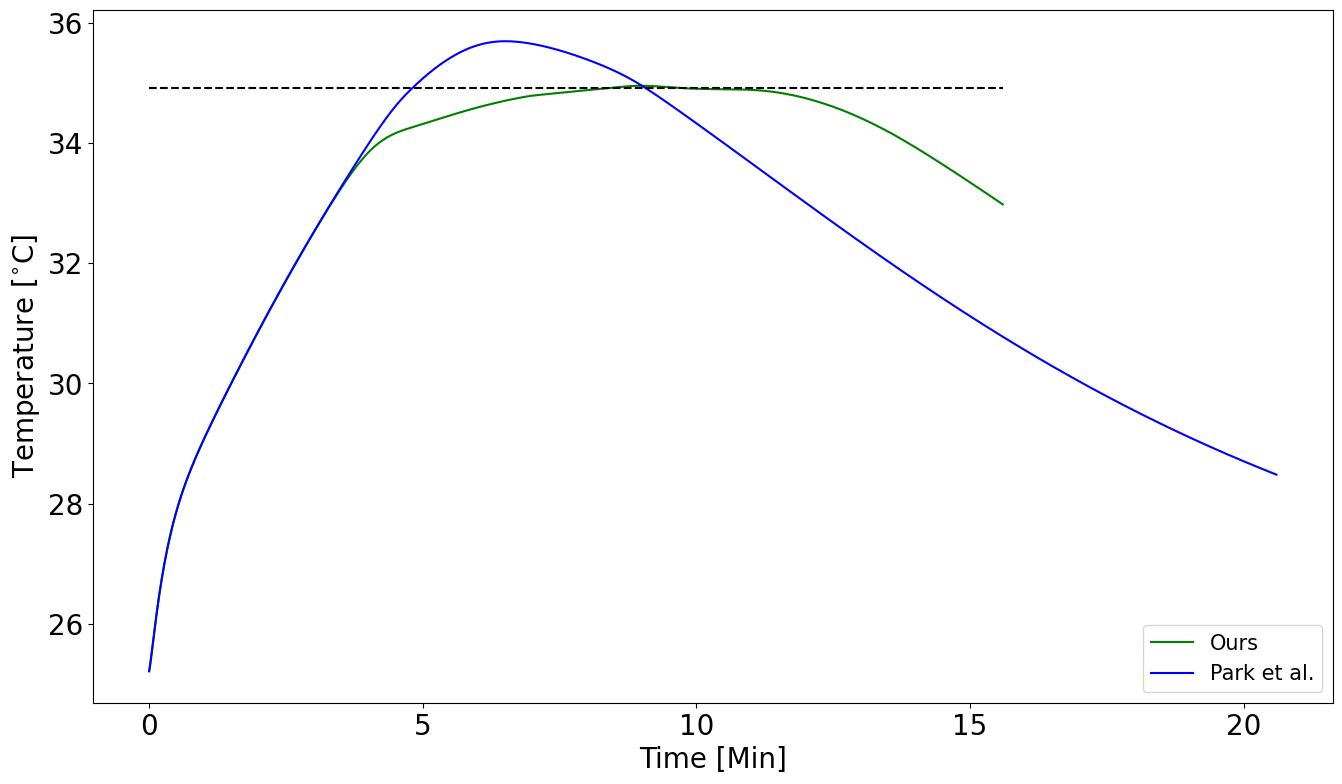}
        \label{fig:TEMP2}
        \centerline{(a)}\medskip
    \end{minipage}

    \begin{minipage}[b]{0.45\textwidth}
        \centering
        \includegraphics[width=\textwidth]{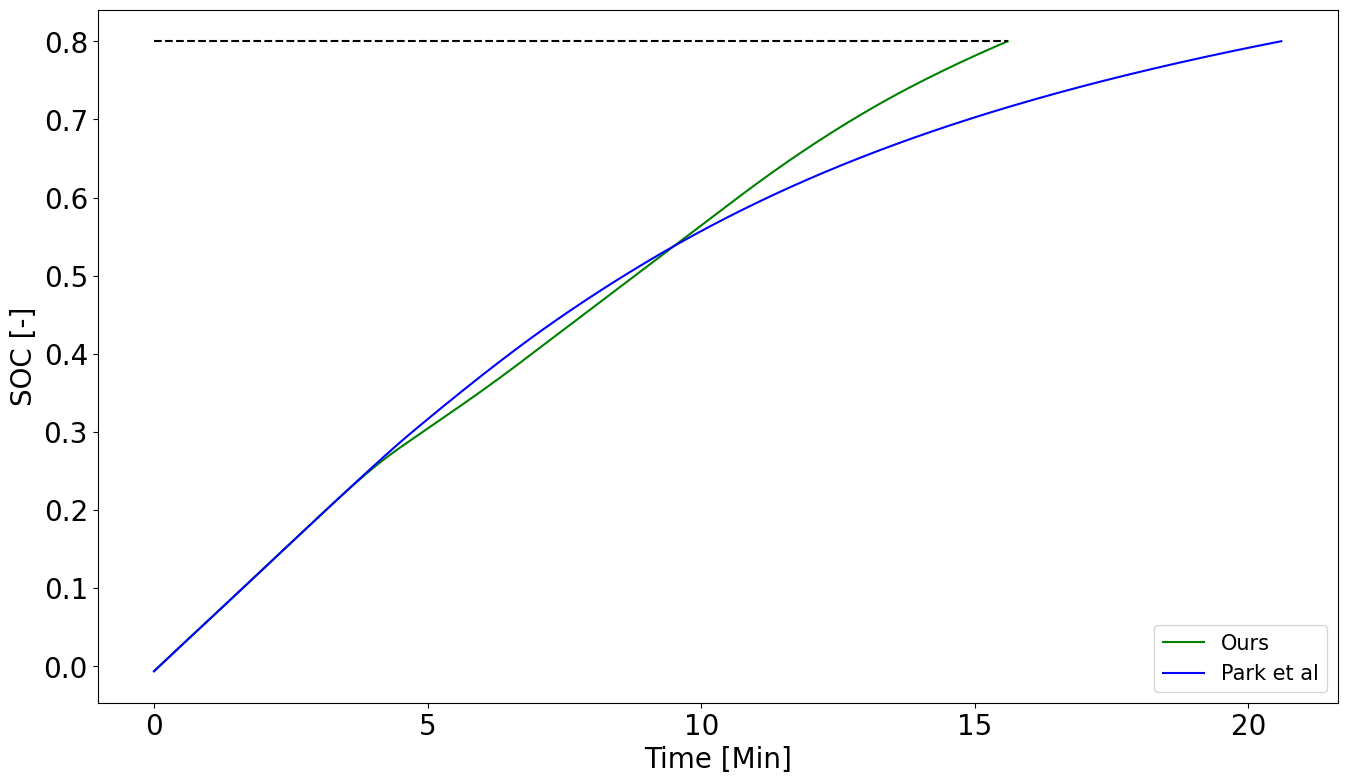}
        \label{fig:SoC2}
        \centerline{(b)}\medskip
    \end{minipage}
    \begin{minipage}[b]{0.45\textwidth}
        \centering
        \includegraphics[width=\textwidth]{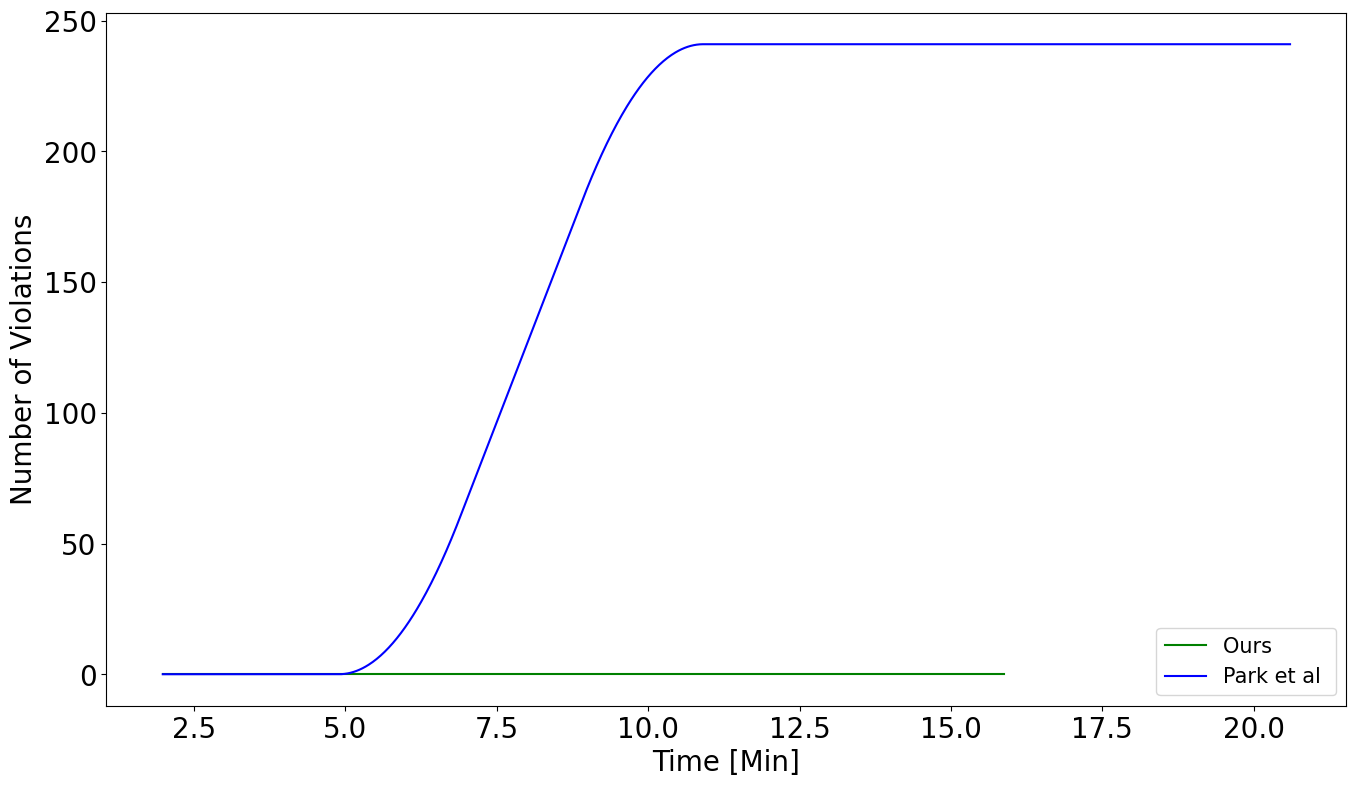}
        \label{fig:VIOALL2}
        \centerline{(c)}\medskip
    \end{minipage}
    \caption{Evaluation of proposed policy on configuration 2 with Initial Temperature=293K and Initial Voltage=2.2V (a)Temperature (b)SoC (c)Violations}
    \label{fig:Config2}
\end{figure}

The plot \ref{fig:config 1}(a) depicts the effect of the action taken by the agent on the battery's temperature. At every time step, we employ a safety network that perturbs the original action to make sure it is a safe one, unlike the Park et al.\cite{park2022deep} approach. The effect of this perturbation on the temperature can be seen in the figure. Our approach has shown that the temperature of the battery does not go beyond the maximum safe temperature of 35 degrees Celsius.
Further, it is evident that our policy speeds up battery charging—it only takes \textbf{15 minutes} to reach 80\% SoC while adhering to safety regulations. However, the baseline takes \textbf{20 minutes} and the battery also exceeds the safe temperature.
Figure \ref{fig:config 1}(b) illustrates a crucial aspect of battery charging: SoC. In comparison to the other policy, our battery SoC follows a linear trend. This linear trend in SoC while charging indicates that the battery is charging fast at an almost constant current and voltage. As a result, it effectively demonstrates that our approach provides fast charging while adhering to safety constraints, i.e. keeping the battery within a safe operating area.

The second configuration on which we have tested our learned policy is with an initial temperature of 293 Kelvin and an initial voltage of 2.2V.
Figure \ref{fig:Config2} illustrate the plots on which our policy has been evaluated along with the baseline policy. In Figure \ref{fig:Config2}(a), it can be seen that the action taken by our learned policy after the 5th minute ensures that the temperature safety constraint is not violated. As a result, it can be seen in figure \ref{fig:Config2}(b) 
the battery achieves 80\% SoC level in roughly \textbf{16 minutes}, whereas the baseline policy takes more than \textbf{20 minutes}. Furthermore, the baseline policy has temperature safety violations, as opposed to the proposed safety layer DDPG policy. 

To better quantify the violations, we show the number of violations in both configurations. The number of violations is estimated in the same way as in equations \ref{eqntempviol} and \ref{eqnvoltvio}. The plots in \ref{fig:config 1}(c) and \ref{fig:Config2}(c) show that our policy does not break the safety requirements in both setups, but the baseline DDPG violates them multiple times.

Based on these results, it can be concluded that our agent's policy works well to ensure safety even in settings where the chances of constraint violations are higher while charging faster. As previously stated, this safety maintenance extends battery life while also preventing hazardous conditions such as thermal runaway or fire dangers.
\section{Conclusion}
\label{sec:con}
In this paper, we discuss a method to ensure fast charging while adhering to safety constraints during the charging of Li-ion batteries using the Reinforcement Learning framework. We introduce a safety layer that consists of a safety network to give a safety signal based on the state of the battery. Our method uses quadratic programming problem techniques to perturb the original action based on the safety signal and output a safe action. By leveraging the safety layer for the action perturbations for safe exploration, we exhibit better performance when compared to the baseline by managing the trade-off between fast charging and safety. Our empirical results showcase that our method achieves SoC faster, with minimal violations of safety constraints. Further, we have trained our policy with three different battery configurations. This ensures that a more generalized policy is learned by the agent. Moreover, the learned policy has been tested in different battery configurations with different starting conditions for validation. Our Validation results show that the learned policy is robust and can be applied to dynamic environments.
\bibliographystyle{ACM-Reference-Format.bst}
\bibliography{software}

\end{document}